\documentclass[epj, twocolumn]{webofc}
\usepackage[varg]{txfonts}   % Web of Conferences font

\usepackage[english]{babel}
\usepackage{enumerate}
\usepackage{amsmath,amssymb}
\usepackage{letltxmacro}			% \let for commands with optional arguments
\usepackage{xkeyval}				% programming: <key> = <value>
\usepackage{xargs}					% programming: better newcommand
\usepackage{xifthen}				% programming: provides \isempty test
\usepackage{slashed}
\usepackage{bm}					% bold math
\usepackage{dsfont}
\usepackage[cmyk,usenames,dvipsnames,svgnames,table,x11names]{xcolor}
\usepackage{tikz}
\usepackage{tikz-3dplot}
\usetikzlibrary{calc,matrix, shapes,arrows,decorations.pathmorphing,decorations.markings,decorations.pathreplacing,trees}
\usepackage[pdftex, unicode=true, pdfnewwindow=true, colorlinks=true, linkcolor=black, urlcolor=black, citecolor=black, hyperindex, pdfauthor={Frederik F. Van der Veken}, pdftitle={Working with Piecewise Linear Wilson Lines}, bookmarksopen=true, bookmarksopenlevel=0]{hyperref}
\makeatletter
	\let\oldr@@t\r@@t
	\def\r@@t#1#2{%
	\setbox0=\hbox{$\oldr@@t#1{#2\,}$}\dimen0=\ht0
	\advance\dimen0-0.2\ht0
	\setbox2=\hbox{\vrule height\ht0 depth -\dimen0}%
	{\box0\lower0.4pt\box2}}
	\LetLtxMacro{\oldsqrt}{\sqrt}
	\renewcommand*{\sqrt}[2][\ ]{\oldsqrt[#1]{#2}}
\makeatother

\newcommandx*{\1}[2][1,2]{\ensuremath{
		\ifthenelse{\isempty{#1} \AND \isempty{#1}}{\mathds{1}}{
			\ifthenelse{\isempty{#2}}
				{\mathds{1}_{#1}}
				{\mathds{1}_{{#1}\times {#2}}}
}
}}

\renewcommand*{\bar}{\overline}

\newcommand{\Not}{\stackrel{\text{\tiny N}}{=}}							% for notational definitions
\renewcommand{\exp}[1]{\ensuremath{\text{e}^{#1}}}
\renewcommand*{\i}{\text{i}\hspace*{1pt}}

\newcommandx*{\wilson}[3][3]{\ensuremath{\mathcal{U}_{(#1 \, ; \,  #2)}^{#3}}}

\newlength{\nplength}
\newcommandx{\negphantom}[1]{\settowidth{\nplength}{#1}\hspace{-\nplength}}

\renewcommandx*{\vec}[3][2, 3]{{\ensuremath{\mathbf{\bm{#1}}_\mathrm{#2}^{\mathrm{#3}}}}}
\newcommandx*{\Trace}[1][1]{\ensuremath{
		\:\textrm{Tr}\ifthenelse{\isempty{#1}}{\,}{\!\left( #1 \right)}
}}
\newcommandx*{\trace}[1][1]{\ensuremath{
		\:\textrm{tr}\ifthenelse{\isempty{#1}}{\,}{\!\left( #1 \right)}
}}

\newcommandx*{\diracdelta}[2][1]{\ensuremath{
		\ifthenelse{\isempty{#1}}
		{\,\delta\!\left({#2}\right)}
		{\,\delta^{\IfInteger{#1}{({#1})}{{#1}}}\!\left({#2}\right)}
}}
\newcommand{\heavisidetheta}[1]{\, \theta\!\left( #1 \right)}

\newcommandx*{\Int}[4][1, 2, 4={0pt}, usedefault, addprefix=\global]{
	\ensuremath{\int\limits_{\:\!{#1}}^{\:\!{#2}}{\!\!}\hspace{#4} {\protect#3}\,\,}
}

\newcommandx*{\Dif}[2][1]{\ensuremath{{\textrm{d}}^{{#1}} {#2}\,}}
\newcommandx*{\dif}[2][1]{\ensuremath{\partial^{{#1}} {#2}\,}}
\newcommandx*{\DDif}[2][1]{\ensuremath{{\mathcal{D}}^{{#1}} {#2}\,}}
\newcommandx*{\Diff}[8][1, 4, 5, 6, 7, 8]{\ensuremath{
		\ifthenelse{\isempty{#4} \AND \isempty{#5}}
			{\frac{\Dif[#1]{#2}}{{\Dif{#3}}^{{#1}}}}
			{\frac{\Dif[#1]{#2}}{
				\ifthenelse{\isempty{#4}} {\Dif{#3}} {\left(\Dif{#3}\right)^{#4}}
				\ifthenelse{\isempty{#5}} {} { \ifthenelse{\isempty{#6}} {\Dif{#5}} {\left(\Dif{#5}\right)^{#6}}}
				\ifthenelse{\isempty{#7}} {} { \ifthenelse{\isempty{#8}}{\Dif{#7}}{\left(\Dif{#7}\right)^{#8}}}
}}}}
\newcommandx*{\diff}[8][1, 4, 5, 6, 7, 8]{\ensuremath{
		\ifthenelse{\isempty{#4} \AND \isempty{#5}}
			{\frac{\dif[#1]{#2}}{{\dif{#3}}^{{#1}}}}
			{\frac{\dif[#1]{#2}}{
				\ifthenelse{\isempty{#4}} {\dif{#3}} {\left(\dif{#3}\right)^{#4}}
				\ifthenelse{\isempty{#5}} {} { \ifthenelse{\isempty{#6}} {\dif{#5}} {\left(\dif{#5}\right)^{#6}}}
				\ifthenelse{\isempty{#7}} {} { \ifthenelse{\isempty{#8}}{\dif{#7}}{\left(\dif{#7}\right)^{#8}}}
}}}}

\newcommandx*{\braket}[3][1,2,3]{
	\ifthenelse{\isempty{#2} \AND \isempty{#3}}
	{\ifthenelse{\isempty{#1}}  {\opm{!! empty braket used !!}}  {\ensuremath{\left< {#1} \right>} } }
	{\ifthenelse{\isempty{#3}}
			{\ensuremath{\left< {#1} \vphantom{{#2}}  \:\! \right| \! \! \! \; \left. {#2} \vphantom{{#1}} \right>} }
			{\ensuremath{\left< {#1} \vphantom{{#3}} \right| #2 \left| {#3} \vphantom{{#1}} \right>} }
}}

\newenvironment{subalign}[1][]{
	\subequations
	\ifthenelse{\isempty{#1}}{}{\label{#1}}
	\align
}{
	\endalign
	\endsubequations
}

\definecolor{geel}{cmyk}{0.,0.075,1.,0.2}
\definecolor{blauw}{cmyk}{1.,0.55,0.05,0.05}
\definecolor{groen}{cmyk}{0.6,0.15,1.,0.1}
\definecolor{rood}{cmyk}{0.1,1.,0.1,0.3}
\newcommand{\photoncolour}{geel}
\newcommand{\gluoncolour}{groen}
\newcommand{\fermioncolour}{geel}
\newcommand{\darkfermioncolour}{geel!90!blauw!80!black}
\newcommand{\wilsoncolour}{blauw}
\newcommand{\eikonalcolour}{rood}
\newcommand{\bloboutercolour}{gray!80!black}
\newcommand{\blobinnercolour}{gray!20}
\newcommand{\pdfoutercolour}{geel!80!groen!80!blauw}
\newcommand{\pdfinnercolour}{geel!60!groen!80!blauw!20!white}

\pgfdeclaredecoration{complete sines}{initial}{				% better photon decoration
	\state{initial}[
		width=+0pt,
		next state=sine,
		persistent precomputation={
			\pgfmathsetmacro\matchinglength{
				\pgfdecoratedinputsegmentlength / int(\pgfdecoratedinputsegmentlength/\pgfdecorationsegmentlength)
			}
			\setlength{\pgfdecorationsegmentlength}{\matchinglength pt}
        		}
	] {}
	\state{sine}[width=\pgfdecorationsegmentlength]{
		\pgfpathsine{\pgfpoint{0.25\pgfdecorationsegmentlength}{0.5\pgfdecorationsegmentamplitude}}
		\pgfpathcosine{\pgfpoint{0.25\pgfdecorationsegmentlength}{-0.5\pgfdecorationsegmentamplitude}}
		\pgfpathsine{\pgfpoint{0.25\pgfdecorationsegmentlength}{-0.5\pgfdecorationsegmentamplitude}}
		\pgfpathcosine{\pgfpoint{0.25\pgfdecorationsegmentlength}{0.5\pgfdecorationsegmentamplitude}}
	}
	\state{final}{}
}
\pgfdeclaredecoration{integralshape}{integralshape}{		% for cut lines
	\state{integralshape}[width=+\pgfdecoratedremainingdistance,next state=final] {
		\pgfpathmoveto{\pgfpoint{0}{0.042*\pgfdecoratedpathlength*\pgfdecorationsegmentamplitude}}
		\pgfpathcurveto
			{\pgfpoint{-0.01*\pgfdecoratedpathlength*\pgfdecorationsegmentamplitude}{0.02*\pgfdecoratedpathlength*\pgfdecorationsegmentamplitude}}
			{\pgfpoint{\pgfdecoratedpathlength*(0.5-0.18*\pgfdecorationsegmentamplitude)}{0.18*tan(\pgfdecorationsegmentangle)*\pgfdecorationsegmentamplitude*\pgfdecoratedpathlength}}
			{\pgfpoint{\pgfdecoratedpathlength*0.5}{0}}
		\pgfpathcurveto
			{\pgfpoint{\pgfdecoratedpathlength*(0.5+0.18*\pgfdecorationsegmentamplitude)}{-0.18*tan(\pgfdecorationsegmentangle)*\pgfdecorationsegmentamplitude*\pgfdecoratedpathlength}}
			{\pgfpoint{\pgfdecoratedpathlength*(1+0.01*\pgfdecorationsegmentamplitude)}{-0.02*\pgfdecoratedpathlength*\pgfdecorationsegmentamplitude}}
			{\pgfpoint{\pgfdecoratedpathlength}{-0.042*\pgfdecoratedpathlength*\pgfdecorationsegmentamplitude}}
	}
	\state{final}{}
}

\def\myscale{1}
\tikzset{
photon/.style={decorate, decoration={snake,amplitude=\myscale*3pt, segment length=\myscale*7pt}, draw=\photoncolour, line width=\myscale*1.4pt},
gluon/.style={decorate, draw=\gluoncolour, decoration={coil,amplitude=\myscale*3pt, segment length=\myscale*4pt},  line width=\myscale*1.2pt},
quark/.style={draw=\fermioncolour, postaction={decorate},
	decoration={markings,mark=at position .5*\pgfdecoratedpathlength+sqrt(\myscale)*5.5pt with {\arrow[\fermioncolour]{latex}}},  line width=\myscale*1.6pt},		%arrowhead can also be ' > ', it is a matter of taste
quarknoarrow/.style={draw=\fermioncolour, line width=\myscale*1.6pt},
darkquark/.style={draw=\darkfermioncolour, postaction={decorate},
	decoration={markings,mark=at position .5*\pgfdecoratedpathlength+sqrt(\myscale)*5.5pt with {\arrow[\darkfermioncolour]{latex}}},  line width=\myscale*1.6pt},
eikonal/.style={double, double distance=\myscale*2.5pt, draw=\eikonalcolour, postaction={decorate},
	decoration={markings,mark=at position .5*\pgfdecoratedpathlength+sqrt(\myscale)*7.5pt with {\arrow[\eikonalcolour]{angle 60}}}, line width=\myscale*1.2pt},
wilson/.style={double, double distance=\myscale*2.5pt, line width=\myscale*1.2pt, draw=\wilsoncolour},
blob/.style={draw=\bloboutercolour, fill=\blobinnercolour, line width=\myscale*1.2pt},
pdf/.style={draw=\pdfoutercolour, fill=\pdfinnercolour, line width=\myscale*1.2pt},
photontext/.style={\photoncolour!80!black},
gluontext/.style={\gluoncolour!80!black},
quarktext/.style={\fermioncolour!80!black},
wilsontext/.style={\wilsoncolour!80!black},
eikonaltext/.style={\eikonalcolour!80!black},
blobtext/.style={\bloboutercolour!80!black},
pdftext/.style={\pdfoutercolour!80!black},
accolade/.style={gray,decorate, decoration={brace,amplitude=\myscale*5pt}, line width=\myscale*1.2pt},
hide/.style={ultra thick, white},
finalstatecut/.style={decorate, decoration={integralshape,amplitude=(1/\myscale)*1.5pt,angle=2}, line width=\myscale*1.2pt},
wilsonarrow/.style={postaction={decorate}, decoration={markings,mark=at position .75 with {\arrow[\wilsoncolour]{angle 60}}}},
wilsonarrowreversed/.style={postaction={decorate}, decoration={markings,mark=at position .3 with {\arrowreversed[\wilsoncolour]{angle 60}}}},
wilsonarrow2/.style={postaction={decorate}, decoration={markings,mark=at position 1. with {\arrow[\wilsoncolour]{angle 60}}}},
wilsonarrowreversed2/.style={postaction={decorate}, decoration={markings,mark=at position 0 with {\arrowreversed[\wilsoncolour]{angle 60}}}}
}
\newenvironment{tikzfigure}[2]{
	\def\myscale{#1}
	\begin{tikzpicture}[baseline= {($(current bounding box.base)-(0pt,#2)$)},scale=\myscale]
}
{
	\end{tikzpicture}
}

\woctitle{Transversity 2014}

\begin{document}

\title{Piecewise linear Wilson lines}
\author{Frederik F. Van der Veken\inst{1}\fnsep\thanks{\email{frederik.vanderveken@ua.ac.be}}
}

\institute{Department of Physics, University of Antwerp, Groenenborgerlaan 171, 2020 Antwerpen, Belgium}

\abstract{
  Wilson lines, being comparators that render non-local operator products gauge invariant, are extensively used in QCD calculations, especially in small-$x$ calculations, calculations concerning validation of factorisation schemes and in calculations for constructing or modelling parton density functions. We develop an algorithm to express piecewise path ordered exponentials as path ordered integrals over the separate segments, and apply it on linear segments, reducing the number of diagrams needed to be calculated. We show how different linear path topologies can be related using their colour structure. This framework allows one to easily switch results between different Wilson line structures, which is especially useful when testing different structures against each other, e.g.\@ when checking universality properties of non-perturbative objects.
}
%\label{sec-1}
\maketitle

\section{Introduction}
In the so-called \emph{eikonal approximation},  a moving quark is considered to emit only soft and collinear radiation, which can be resummed into a Wilson line. One case where we could use the eikonal approximation, is for a quark in a dense medium (see e.g.\@ \cite{Blaizot:2012fh} and \cite{Iancu:2013dta}, where in the latter the medium is reduced into a shockwave).

A Wilson line on a path with endpoints $a^\mu$ and $b^\mu$ transforms as
\begin{equation}\label{eq: gaugetransform}
  \wilson{b}{a} \rightarrow
    \exp{\i \, g\alpha(b)} \wilson{b}{a} \, \exp{-\i\, g \alpha(a)},
\end{equation}
which can be utilised to make bilocal operator products gauge-invariant. E.g.\@ in the TMD approach, the operator definition for a transverse momentum dependent parton density function contains a bilocal product of quark field operators $\bar{\psi}(z)\psi(0)$ \cite{Collins:2011zzd,Echevarria:2012js}. One then inserts a Wilson line (also called a \emph{gauge link} in this approach) such that the resulting factor $\bar{\psi}(z)\,\wilson{z}{0}\psi(0)$ is gauge invariant. As the gauge transformation of the Wilson line only depends on its endpoints, there is some freedom on the choice of the path. The correct path will be constructed by eikonalising the quark fields and identifying the appropriate gluon radiation \cite{Belitsky:2002sm,Boer:2003cm,Cherednikov:2008ua}.

Other applications of Wilson lines include the calculation of soft factors \cite{GarciaEchevarria:2011rb,Becher:2012za,Idilbi:2007ff,Chiu:2009yx}, the study of jet quenching \cite{Cherednikov:2013pba}, a recast of QCD in loop space where the geometric evolution of rectangular loops can be related to its energy evolution \cite{Mertens:2014hma, Cherednikov:2012qq,VanderVeken:2014lka,Mertens:2014lla,Mertens:2014mia}, etc.

A general Wilson line is an exponential of gauge fields along a path $\mathcal{C}$, defined as follows:
\begin{equation}\label{eq: Wilsondef}
  \mathcal{U} =
    \mathcal{P}\,\exp{\i g \int_\mathcal{C} \Dif{z^\mu} A_\mu (z)}.
\end{equation}
Because the gauge fields are non-Abelian, i.e.\@ $A_\mu = t^a A^a_\mu$, they have to be \emph{path ordered}, denoted by the symbol $\mathcal{P}$ in \eqref{eq: Wilsondef}, to avoid ambiguities. The fields are ordered such that the fields first on the path are written leftmost. After making a Fourier transform the path content is fully described by the following integrals:
\begin{equation} \label{eq: PathIntegrals}
  I_n =
    \frac{1}{n!}\; 
    \mathcal{P} \!\!\int\! \Dif{\lambda_1}\cdots \Dif{\lambda_n}
    \left(z_1^{\mu_1}\right)' \cdots \left(z_n^{\mu_n}\right)'
    \exp{-\i\prod\limits^n k_i\cdot z_i},
\end{equation}
such that the $n$-th order term of the Wilson line expansion is given by
\begin{equation}
  \mathcal{U}^n =
    \left(\i g\right)^n \! \Int{\frac{\Dif[4]{k_1}}{(2\pi)^4}\cdots\frac{\Dif[4]{k_n}}{(2\pi)^4}}
      A_{\mu_n}(k_n)\!\cdots\! A_{\mu_1}(k_1) \; I_n.
\end{equation}

\section{Piecewise Path Ordered Integrals}\label{sec: piecewiseintegrals}
We will investigate piecewise functions consisting of $M$ continuously differentiable segments (in particular, each segment should be defined over an interval that is not a single point). The path-ordered integral over one segment is simply
\begin{equation}
  S_2^J =
    \Int[a_J][a_{J+1}]{\Dif{x_1}} \!\!\!\!\Int[a_J][x_1]{\Dif{x_2}} \dotsm
    \Int[a_J][x_{n-1}]{\Dif{x_n}}
     f^J(x_1) f^J(x_2)\dotsm f^J{x_n},
\end{equation}
Such that we can express the path-integral over the full path in function of the segment integrals as
\begin{equation}
  I_n = \label{eq: PiecewiseIntegral}
    \sum_{i=1}^n \left[
      \left(\prod_{j=1}^i\sum\limits_{J_j=i-j+1}^{J_{j-1}-1}\right)_{J_0-1=M}
      \left(
        \begin{array}{c}
          \text{All }\prod\limits_{j=1}^i S_{l_j}(J_j) \\
          \text{where } \sum\limits_{j=1}^i l_j =n
        \end{array}
      \right)
    \right].
\end{equation}
E.g. the third order integral is given by
\begin{equation}
  I_3 =\label{eq: SegmentOrder3}
    \sum\limits_{J=1}^{M}S_3^J +
    \sum\limits_{J=2}^{M}\sum\limits_{K=1}^{J-1}
      \bigg[ S_1^J S_2^K + S_2^J S_1^K \bigg] +
    \sum\limits_{J=3}^{M}\sum\limits_{K=2}^{J-1}\sum\limits_{L=1}^{K-1}
      S_1^J S_1^K S_1^L, \\ \nonumber
\end{equation}
It is also possible to give a recursive definition:
\begin{equation}\label{eq: RecursiveDef}
  I_n(M) = \sum\limits_{J=1}^M S_n^J + \sum\limits_{J=2}^M \sum\limits_{i=1}^{n-1} S_i^J \;I_{n-i}(J-1).
\end{equation}
Eqs.\@ \eqref{eq: PiecewiseIntegral} and \eqref{eq: RecursiveDef} literally translate to a Wilson line; just replace every $S$ with an $U$, for instance:
\begin{equation*}
  \mathcal{U}_3 =\!
    \sum\limits_{J=1}^{M}\mathcal{U}_3^J +
    2\!\sum\limits_{J=2}^{M}\sum\limits_{K=1}^{J-1}
      \!\mathcal{U}_{(1}^J \mathcal{U}_{2)}^K +\!
    \sum\limits_{J=3}^{M}\sum\limits_{K=2}^{J-1}\sum\limits_{L=1}^{K-1}
      \!\mathcal{U}_1^J \mathcal{U}_1^K \mathcal{U}_1^L,
\end{equation*}
where
\begin{equation*}
  \mathcal{U}_n^J =
    \left(\frac{\i g}{16\pi^4}\right)^{\!n}\!
    \Int{\Dif[4]{k_1}\cdots \Dif[4]{k_n}} A_{\mu_1}(k_1) \cdots A_{\mu_n}(k_n) \;
    S_n^J.
\end{equation*}
The physical interpretation of the $n$-th order formula is a collection of all possible diagrams for $n$-gluon radiation from a $M$-segment Wilson line.

Consider now the product of e.g.\@ three Wilson lines, labelled $\mathcal{U}^A$, $\mathcal{U}^B$ and $\mathcal{U}^C$. Expanding the exponentials and collecting terms of the same order in $g$ we get:
\begin{align*}
  \mathcal{U}^A &\mathcal{U}^B \mathcal{U}^C =
    1 + \left(\mathcal{U}^A_1 + \mathcal{U}^B_1 + \mathcal{U}^C_1\right) \\
  &\,+
    \left(\mathcal{U}^A_1 \mathcal{U}^B_1 + \mathcal{U}^A_1 \mathcal{U}^C_1 + \mathcal{U}^B_1 \mathcal{U}^C_1+ \mathcal{U}^A_2 + \mathcal{U}^B_2 + \mathcal{U}^C_2\right) \\
  &\,+
    \left( \mathcal{U}^A_1\mathcal{U}^B_1\mathcal{U}^C_1 + \mathcal{U}^A_1\mathcal{U}^B_2 +
    \mathcal{U}^A_1\mathcal{U}^C_2 + \mathcal{U}^B_1\mathcal{U}^C_2 + \mathcal{U}^A_2 \mathcal{U}^B_1 \right. \\
  &\,\phantom{++}+\left.
    \mathcal{U}^A_2 \mathcal{U}^C_1 + \mathcal{U}^B_2 \mathcal{U}^C_1 +
    \mathcal{U}^A_3 + \mathcal{U}^B_3 + \mathcal{U}^C_3
    \right) + \ldots,
\end{align*}
which equals, up to third order, the sum of the first, second and third order full integrals. In other words, we can equate a product of Wilson lines to one Wilson line consisting of several segments. The proof easily generalises to all orders. Note that the order of the segments is reversed w.r.t.\@ the order of the product (because we read the lines from right to left), e.g.\@ the product $\mathcal{U}^A \mathcal{U}^B \mathcal{U}^C$ is a line $\mathcal{U}^{CBA}$, with first segment $C$, second segment $B$ and last segment $A$.

\section{Linear Path Segments}
The results from the former section are general results, i.e.\@ valid for any path. Let us now turn our focus towards linear paths only, as these are the most commonly used in literature. For every segment there exist four possible path structures: it can be a finite segment connecting two points, it can be a segment connecting $\pm\infty$ and a point $r^\mu$, or it can be a fully infinite line connecting $-\infty$ with $+\infty$. \\

We start by considering a line from a point $a^\mu$ to $+\infty$, along a direction $\hat{n}^\mu$. Such a path can be parameterised as
\begin{equation}
  z^\mu = a^\mu + \lambda \, \hat{n}^\mu \quad \lambda = 0 \ldots \infty.
\end{equation}
Using \eqref{eq: PathIntegrals}, it is straightforward to calculate the segment integrals for a lower bound Wilson line:
\begin{equation}
  I_n^\text{l.b.} =  \label{eq: lowerbound}
    \hat{n}^{\mu_1}\cdots \hat{n}^{\mu_n} \,
    \exp{\i a\cdot \sum\limits_j k_j} \,
    \prod\limits_{j=1}^n
    \frac{\i}{\hat{n}\!\cdot\! \sum\limits_{l=j}^n k_l + \i \eta}.
\end{equation}
The $\i\eta$ are convergence terms added to regularise the exponent.We can reconstruct this result using the following Feynman rules:
\begin{figure*}[t!]
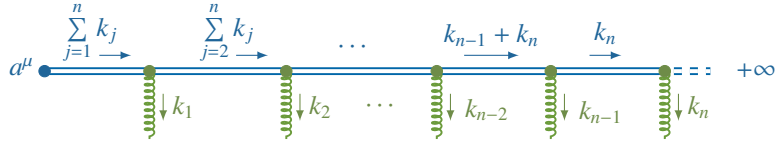

\centering
\begin{tikzfigure}{0.6}{0}
  \draw[wilson] (1.8,0) -- (15.3,0);
  \draw[wilson,dashed] (15.5,0) -- (16.3,0);
  \node[wilsontext] at (17.3,0.03) {$+\infty$};
  \filldraw[wilsontext] (1.7,0) circle(0.125);
  \node[wilsontext] at (1.2,0.1) {$a^\mu$};
  
  \draw[gluon] (4,0) -- (4,-1.5);
  \filldraw[gluontext] (4,0) circle (0.125);
  \draw[-latex, gluontext] (4.3,-0.5) -- (4.3, -1.1);
  \node[gluontext] at (4.75,-0.8) {$k_1$};
  \draw[-latex,wilsontext] (2.9,0.35) -- (3.6,0.35);
  \node[wilsontext] at (2.7,0.85) {$\sum\limits_{j=1}^{n} k_j $};
  \begin{scope}[shift={(3,0)}]
    \draw[gluon] (4,0) -- (4,-1.5);
    \filldraw[gluontext] (4,0) circle (0.125);
    \draw[-latex, gluontext] (4.3,-0.5) -- (4.3, -1.1);
    \node[gluontext] at (4.75,-0.8) {$k_2$};
    \draw[-latex,wilsontext] (2.9,0.35) -- (3.6,0.35);
    \node[wilsontext] at (2.7,0.85) {$\sum\limits_{j=2}^{n} k_j $};
  \end{scope}
  \node[wilsontext] at (8.5,0.5) {$\cdots$};
  \node[gluontext] at (9.1,-0.8) {$\cdots$};
  \begin{scope}[shift={(6.3,0)}]
    \draw[gluon] (4,0) -- (4,-1.5);
    \filldraw[gluontext] (4,0) circle (0.125);
    \draw[-latex, gluontext] (4.3,-0.5) -- (4.3, -1.1);
    \node[gluontext] at (5.1,-0.85) {$k_{n-2}$};
  \end{scope}
  \begin{scope}[shift={(8.8,0)}]
    \draw[gluon] (4,0) -- (4,-1.5);
    \filldraw[gluontext] (4,0) circle (0.125);
    \draw[-latex, gluontext] (4.3,-0.5) -- (4.3, -1.1);
    \node[gluontext] at (5.1,-0.9) {$k_{n-1}$};
    \draw[-latex,wilsontext] (2.1,0.35) -- (3.3,0.35);
    \node[wilsontext] at (2.7,0.8) {$ k_{n-1}+k_n $};
  \end{scope}
  \begin{scope}[shift={(11.3,0)}]
    \draw[gluon] (4,0) -- (4,-1.5);
    \filldraw[gluontext] (4,0) circle (0.125);
    \draw[-latex, gluontext] (4.3,-0.5) -- (4.3, -1.1);
    \node[gluontext] at (4.75,-0.8) {$k_n$};
    \draw[-latex,wilsontext] (2.35,0.35) -- (3.05,0.35);
    \node[wilsontext] at (2.7,0.85) {$k_n $};
  \end{scope}
\end{tikzfigure}
\caption{$n$-gluon radiation for a Wilson line going from $a^\mu$ to $+\infty$.}
\label{fig: lowerbound}
\end{figure*}
\begin{subalign}[eq: Feynman rules]
  \emph{Propagator:} & \hspace*{1.72em} \;
    \begin{tikzfigure}{0.6}{0}
      \draw[wilson] (0,0) -- (2,0);
      \node[wilsontext] at (1,0.65) {$k$};
      \draw[-latex, wilsontext] (0.7,0.25) -- (1.3,0.25);
    \end{tikzfigure} &&\hspace{-1em}
    = \frac{\i}{\hat{n} \!\cdot\! k + \i\eta} \\
  \emph{External point:} & \; \label{eq: FRextpoint}
    \begin{tikzfigure}{0.6}{0.1}
      \draw[wilson] (0,0) -- (2,0);
      \node[wilsontext] at (1,0.65) {$k$};
      \draw[-latex,wilsontext] (0.7,0.25) -- (1.3,0.25);
      \filldraw[wilsontext] (0,0) circle(0.125);
      \node[wilsontext] at (-0.6,0.1) {$a^\mu$};
    \end{tikzfigure} &&\hspace{-1em}
    = e^{\i a \cdot k} \\[1em]
  \emph{Line to infinity:} & \hspace*{1.72em} \;
    \begin{tikzfigure}{0.6}{2pt}
      \draw[wilson,dashed] (0,0) -- (2,0);
      \node[wilsontext] at (2.9,0.03) {$+\infty$};
    \end{tikzfigure} &&\hspace{-1em}
    = 1 \\[0.5em]
  \emph{Wilson vertex:} & \hspace*{0.72em} \; \label{eq: FRvertex}
    \begin{tikzfigure}{0.6}{0.3}
      \draw[wilson] (0,0) -- (2,0);
      \draw[gluon] (1,0) -- (1,-1);
      \node[wilsontext] at (-0.3,0) {$j$};
      \node[wilsontext] at (2.3,0) {$i$};
      \filldraw[gluontext] (1,0) circle (0.125);
      \node[gluontext] at (1,-1.25) {$\mu, a$};
      \draw[-latex, gray!80!white] (1.3,-0.3) -- (1.3, -0.8);
      \node[gray!80!white] at (1.6,-0.5) {$k$};
    \end{tikzfigure}&&\hspace{-1em}
    = \i g \, \hat{n}^\mu \left(t^a\right)_{ij}
\end{subalign}
In order to use this Feynman rules in a consistent way, the momenta of the gluons should always be pointing outwards, starting from the external point. As an illustration, the resulting $n$-th order diagram is drawn in figure\@ \ref{fig: lowerbound}. \\

Next we investigate a path that starts at $-\infty$ and goes up to a point $b_\mu$, which we parameterise as
\begin{equation}
  z^\mu = b^\mu  + \hat{n}^\mu \, \lambda \qquad \lambda = -\infty \ldots 0.
\end{equation}
The resulting path integral is almost the same as before:
\begin{equation}\label{eq: upperbound}
  I_n^\text{u.b.} =
    \hat{n}^{\mu_1}\cdots \hat{n}^{\mu_n} \,
    \exp{\i b\cdot \sum\limits_j k_j} \,
    \prod\limits_{j=1}^n
    \frac{-\i}{\hat{n}\!\cdot\! \sum\limits_{l=1}^j k_l - \i \eta}
\end{equation}
which differs from \eqref{eq: lowerbound} only in the accumulation of momenta in the denominators and the sign of the convergence terms. We can use the same Feynman rules if we keep the convention that gluon momenta are radiated outwards from the external point. 
Reversing the path flow is the same as taking the Hermitian conjugate of a Wilson line, which is defined for a line from a point $a^\mu$ to a point $b^\mu$ as:
\begin{equation}
  \mathcal{U}^\dagger = \bar{\mathcal{P}}\,\exp{-\i g \Int[a][b]{\,\Dif{z\cdot A}}},
\end{equation}
where the symbol $\bar{\mathcal{P}}$ denotes anti-path ordering. To be able to treat all segments on an equal basis, we put the fields in the Hermitian conjugate in non-reversed order by reversing the momentum labels. This allows us to relate them to the former results \eqref{eq: lowerbound} and \eqref{eq: upperbound}:
\begin{subalign}[eq: hermitian]
  \wilson{+\infty}{a}[\dagger] &= \wilson{a}{-\infty}\Big|_{\hat{n} \,\rightarrow\,-\hat{n}}\,, \\
  \wilson{b}{-\infty}[\dagger] &= \wilson{+\infty}{b}\Big|_{\hat{n} \,\rightarrow\,-\hat{n}}\,.
\end{subalign}
We will indicate the direction of $\hat{n}$ with a blue arrow on the Wilson line.

We now introduce a shorthand notation to denote the path structure for a Wilson line segment. The first two structures we calculated before are represented by:
\begin{subalign}
  \wilson{+\infty}{a} &\quad \Not \quad 
    \begin{tikzfigure}{0.6}{0.08}
      \draw[wilson, wilsonarrow] (0,0) -- (2,0);
      \filldraw[wilsontext] (0,0) circle(0.125);
    \end{tikzfigure}\,, \\
  \wilson{b}{-\infty} &\quad \Not \quad 
    \begin{tikzfigure}{0.6}{0.08}
      \draw[wilson, wilsonarrow] (0,0) -- (2,0);
      \filldraw[wilsontext] (2,0) circle(0.125);
    \end{tikzfigure}\,.
\end{subalign}
On the other hand, the two structures with reversed $\hat{n}$ are then represented by:
\begin{subalign}
  %\wilson{+\infty}{a}[\dagger] =
  \wilson{a}{+\infty} = \wilson{a}{-\infty}\Big|_{\hat{n}\,\rightarrow\,-\hat{n}} &\quad \Not \quad 
    \begin{tikzfigure}{0.6}{0.08}
      \draw[wilson, wilsonarrowreversed] (0,0) -- (2,0);
      \filldraw[wilsontext] (2,0) circle(0.125);
    \end{tikzfigure}\,, \\
  %\wilson{b}{-\infty}[\dagger] =
  \wilson{-\infty}{b} = \wilson{+\infty}{b}\Big|_{\hat{n}\,\rightarrow\,-\hat{n}} &\quad \Not \quad 
    \begin{tikzfigure}{0.6}{0.08}
      \draw[wilson, wilsonarrowreversed] (0,0) -- (2,0);
      \filldraw[wilsontext] (0,0) circle(0.125);
    \end{tikzfigure}\,.
\end{subalign}
A nice feature of this notation is that we get a ``mirror relation'':
\begin{equation*}
  \big(
  \begin{tikzfigure}{0.6}{0.08}
    \draw[wilson, wilsonarrow] (0,0) -- (2,0);
    \filldraw[wilsontext] (0,0) circle(0.125);
  \end{tikzfigure}
  \big)^\dagger
  = \,
  \begin{tikzfigure}{0.6}{0.08}
    \draw[wilson, wilsonarrowreversed] (0,0) -- (2,0);
    \filldraw[wilsontext] (2,0) circle(0.125);
  \end{tikzfigure}\, ,
  \qquad
  \big(
  \begin{tikzfigure}{0.6}{0.08}
    \draw[wilson, wilsonarrow] (0,0) -- (2,0);
    \filldraw[wilsontext] (2,0) circle(0.125);
  \end{tikzfigure}
  \big)^\dagger
  = \,
  \begin{tikzfigure}{0.6}{0.08}
    \draw[wilson, wilsonarrowreversed] (0,0) -- (2,0);
    \filldraw[wilsontext] (0,0) circle(0.125);
  \end{tikzfigure},
\end{equation*}
which is literally the same as \eqref{eq: hermitian}.
\\

Next we investigate a Wilson line on a finite path, going from a point $a^\mu$ to a point $b^\mu$. We parameterise this as:
\begin{equation}
  z^\mu =
    a^\mu  + \hat{n}^\mu \lambda \qquad \lambda = 0 \, \ldots \, \left\lVert b-a \right\rVert.
\end{equation}
Dropping the factor in front of the integral, we find a recursion relation:
\begin{equation}\label{eq: recursionrelation}
  I_n^\text{fin} {\scriptstyle\left(k_1,\ldots,k_n\right) } =
    \frac{\i}{\hat{n} \!\cdot\! k_1 } \!\! \left(\vphantom{\frac{\i}{k_n}} \!
      I_{n-1}^\text{fin.} {\scriptstyle\left(k_1 + k_2,\ldots,k_n\right)} -
      I_{n-1}^\text{fin.} {\scriptstyle\left(k_2,\ldots,k_n\right)} \!
    \right)
\end{equation}
which we can solve exactly by careful inspection (reintroducing the factor in front):
\begin{multline}\label{eq: finiteResult}
  I_n^\text{fin.} =
    \hat{n}^{\mu_1}\cdots \hat{n}^{\mu_n} \,
    \sum_{m=0}^n \;
    \prod\limits_{j=1}^m \; \frac{\i\, \exp{\i a\cdot k_j}}{\hat{n} \!\cdot\!\sum\limits_{l=j}^m k_l } \;
    \prod\limits_{j=m+1}^n  \; \frac{-\i\, \exp{\i b\cdot k_j} }{\hat{n}\!\cdot\! \!\!\!\!\sum\limits_{l=m+1}^{j}k_l }.
\end{multline}
Using the fact that this kind of chained sum can in general be written as a product of two infinite sums:
\begin{equation*}
  \sum_{n=0}^\infty \sum_{m=0}^nA_m B_{n-m} =
    \left(\sum_{i=0}^\infty A_i \right)\left(\sum_{j=0}^\infty B_j\right),
\end{equation*}
we can transform equation \eqref{eq: finiteResult} into a product of two Wilson lines:
\begin{equation}
  \wilson{b}{a} = \wilson{+\infty}{b}[\dagger]\wilson{+\infty}{a} = \wilson{b}{-\infty}\wilson{a}{-\infty}[\dagger],
\end{equation}
which can be illustrated schematically as:
\begin{align*}
  \begin{tikzfigure}{0.6}{0.08}
    \draw[wilson, wilsonarrow] (0,0) -- (2,0);
    \filldraw[wilsontext] (0,0) circle(0.125);
    \filldraw[wilsontext] (2,0) circle(0.125);
  \end{tikzfigure}
  & \quad = \quad
  \begin{tikzfigure}{0.6}{0.08}
    \draw[wilson, wilsonarrow] (0,0) -- (2,0);
    \filldraw[wilsontext] (0,0) circle(0.125);
  \end{tikzfigure}
  \quad \otimes \quad
  \begin{tikzfigure}{0.6}{0.08}
    \draw[wilson, wilsonarrowreversed] (0,0) -- (2,0);
    \filldraw[wilsontext] (2,0) circle(0.125);
  \end{tikzfigure}
  \\
  & \quad = \quad
  \begin{tikzfigure}{0.6}{0.08}
    \draw[wilson, wilsonarrow] (0,0) -- (2,0);
    \filldraw[wilsontext] (0,0) circle(0.125);
  \end{tikzfigure}
  \quad \otimes \quad
  \big(
  \begin{tikzfigure}{0.6}{0.08}
    \draw[wilson, wilsonarrow] (0,0) -- (2,0);
    \filldraw[wilsontext] (0,0) circle(0.125);
  \end{tikzfigure}
  \big)^\dagger.
\end{align*}

Finally, the last possible path structure for a linear segment is a fully infinite line, going from $-\infty$ to $+\infty$ along a direction $\hat{n}^\mu$ and passing through a point $r^\mu$. Such a path can be parameterised as:
\begin{equation}\label{eq: naiveInfPath}
  z^\mu =
    r^\mu  + \hat{n}^\mu \lambda \qquad \lambda = -\infty \, \ldots \, +\infty.
\end{equation}
Naively, one could think that $I_n^\text{inf.}$ consists of $n-\!1$ integrals that evaluate to the Fourier transform of a Heaviside $\theta$-function, and one integral---the outermost---that evaluates to a Dirac $\delta$-function. In fact, although this is mathematically not correct because it leads to a divergent expression of the $\delta$-function, one can show by regularising the path as in \cite{Kotko:2014aba} that this is indeed the result, as long as the $\delta$-function is used in conjunction with the sifting property (and not by its exponential representation). The result is then
\begin{equation} \label{eq: infResult}
   I_n^\text{inf.} =
    \prod\limits_{1}^{n-1} \; \frac{\i}{\hat{n}\!\cdot\!\sum\limits_j^n k_l +\i\eta} \; 
    2 \pi \, \delta\left(\hat{n}\!\cdot\!\sum\limits_1^n k_j +\i\eta \right).
\end{equation}
We conclude that the correct way to draw an infinite Wilson line, is to put all radiated gluons on one side from the point $r^\mu$, where the line piece connecting the point to the first gluon is a cut propagator having the following Feynman rule:
\begin{align}
  \emph{Cut propagator:} & \hspace*{1.72em} \; \label{eq: FRcutprop}
    \begin{tikzfigure}{0.6}{2pt}
      \draw[wilson] (0,0) -- (2,0);
      \draw[finalstatecut, wilsontext] (1.1,-0.6) -- (1.1,0.6);
      \begin{scope}[shift={(-0.4,0)}]
        \node[wilsontext] at (1,0.65) {$k$};
        \draw[-latex,wilsontext] (0.7,0.25) -- (1.3,0.25);
      \end{scope}
    \end{tikzfigure} &&
    = 2\pi \diracdelta{\hat{n}\!\cdot\! k + \i\eta}
\end{align}

\section{Relating  Different Path Topologies}
Not including the infinite line, we can relate the remaining six path structures to each other. If we choose the following two structures:
\begin{subalign}[eq: basicstructures]
  \hspace*{-1em} \label{eq: basicstructuresa}
  \begin{tikzfigure}{0.6}{2pt}
    \draw[wilson,wilsonarrow] (0,0) -- (2,0);
    \filldraw[wilsontext] (0,0) circle(0.125);
  \end{tikzfigure} &=
    \hat{n}^{\mu_1}\cdots \hat{n}^{\mu_n} \,
    \exp{\i r\cdot \sum\limits_j k_j} \,
    \prod\limits_{j=1}^n
    \frac{\i}{\hat{n}\!\cdot\! \sum\limits_{l=j}^n k_l + \i \eta},
  \\
  \hspace*{-1em} \label{eq: basicstructuresb}
  \begin{tikzfigure}{0.6}{2pt}
    \draw[wilson,wilsonarrowreversed] (0,0) -- (2,0);
    \filldraw[wilsontext] (2,0) circle(0.125);
  \end{tikzfigure} &=
    \hat{n}^{\mu_1}\cdots \hat{n}^{\mu_n} \,
    \exp{\i r\cdot \sum\limits_j k_j} \,
    \prod\limits_{j=1}^n
    \frac{-\i}{\hat{n}\!\cdot\! \sum\limits_{l=1}^j k_l + \i \eta},
\end{subalign}
we can express the remaining four in function of them:
\begin{subalign}
  \hspace*{-0.6em}
  \begin{tikzfigure}{0.6}{2pt}
    \draw[wilson,wilsonarrow] (0,0) -- (2,0);
    \filldraw[wilsontext] (2,0) circle(0.125);
  \end{tikzfigure} &\quad = \quad
  \begin{tikzfigure}{0.6}{2pt}
    \draw[wilson,wilsonarrowreversed] (0,0) -- (2,0);
    \filldraw[wilsontext] (2,0) circle(0.125);
  \end{tikzfigure}
  \;\Big|_{\hat{n}\rightarrow - \hat{n}}
  \, ,\\
  \hspace*{-0.6em}
  \begin{tikzfigure}{0.6}{2pt}
    \draw[wilson,wilsonarrowreversed] (0,0) -- (2,0);
    \filldraw[wilsontext] (0,0) circle(0.125);
  \end{tikzfigure} &\quad = \quad
  \begin{tikzfigure}{0.6}{2pt}
    \draw[wilson,wilsonarrow] (0,0) -- (2,0);
    \filldraw[wilsontext] (0,0) circle(0.125);
  \end{tikzfigure}
  \;\Big|_{\hat{n}\rightarrow - \hat{n}}
  \, ,\\
  \hspace*{-0.6em}
  \begin{tikzfigure}{0.6}{2pt}
    \draw[wilson,wilsonarrow] (0,0) -- (2,0);
    \filldraw[wilsontext] (0,0) circle(0.125);
    \filldraw[wilsontext] (2,0) circle(0.125);
  \end{tikzfigure} &\quad = \quad
  \begin{tikzfigure}{0.6}{2pt}
    \draw[wilson,wilsonarrow] (0,0) -- (2,0);
    \filldraw[wilsontext] (0,0) circle(0.125);
  \end{tikzfigure}
  \;\otimes\;
  \begin{tikzfigure}{0.6}{2pt}
    \draw[wilson,wilsonarrowreversed] (0,0) -- (2,0);
    \filldraw[wilsontext] (2,0) circle(0.125);
  \end{tikzfigure}
  \, ,\\
  \hspace*{-0.6em}
  \begin{tikzfigure}{0.6}{2pt}
    \draw[wilson,wilsonarrowreversed] (0,0) -- (2,0);
    \filldraw[wilsontext] (0,0) circle(0.125);
    \filldraw[wilsontext] (2,0) circle(0.125);
  \end{tikzfigure} &\quad = \quad
  \Big(
  \begin{tikzfigure}{0.6}{2pt}
    \draw[wilson,wilsonarrow] (0,0) -- (2,0);
    \filldraw[wilsontext] (0,0) circle(0.125);
  \end{tikzfigure}
  \;\otimes \;
  \begin{tikzfigure}{0.6}{2pt}
    \draw[wilson,wilsonarrowreversed] (0,0) -- (2,0);
    \filldraw[wilsontext] (2,0) circle(0.125);
  \end{tikzfigure}
  \;\Big|_{\hat{n}\rightarrow - \hat{n}}
  \, .
\end{subalign}
The two first structures aren't fully independent either, as they are related by a sign difference and an interchange of momentum indices:
\begin{equation}
  \begin{tikzfigure}{0.6}{2pt}
    \draw[wilson,wilsonarrowreversed] (0,0) -- (2,0);
    \filldraw[wilsontext] (2,0) circle(0.125);
  \end{tikzfigure} \quad = \; (-)^n \,
  \begin{tikzfigure}{0.6}{2pt}
    \draw[wilson,wilsonarrow] (0,0) -- (2,0);
    \filldraw[wilsontext] (0,0) circle(0.125);
  \end{tikzfigure}
  \;\Big|_{\left(k_1,\ldots,k_n\right) \rightarrow \left(k_n,\ldots,k_1\right)}.
\end{equation}
We can exploit this relation when connecting a Wilson line to a blob. This blob can be constructed from any combination of Feynman diagrams, but cannot contain other Wilson lines.\footnote{If one is interested in interactions between different Wilson lines, it is sufficient to treat the different lines as different segments of one line (as is explained in the end of section \ref{sec: piecewiseintegrals}).}
For the structure given in \eqref{eq: basicstructuresa} this is:
\begin{equation*}
  \begin{tikzfigure}{0.6}{0.75}
    \draw[gluon] (0.5,0) -- (0.5,-2);
    \draw[gluon] (1,0) -- (1,-2);
    \draw[gluon] (2.5,0) -- (2.5,-2);
    \node[gluontext] at (1.75,-0.75) {$\ldots$};
    \draw[wilson, wilsonarrow] (0,0) -- (3,0);
    \filldraw[wilsontext] (0,0) circle(0.125);
    \filldraw[blob] (1.5,-2) circle(1.5 and 0.75);
    \node[blobtext] at (1.5,-2) {$F$};
  \end{tikzfigure}
  \! = \;
    \left(\i g\right)^n t^{a_n}\!\cdots t^{a_1}\! \Int{\mathcal{D}k}
    I_n^\text{l.b.} \, F_{\mu_1\cdots\mu_n}^{a_1\cdots a_n}{ (k_1,\ldots, k_n)},
\end{equation*}
where we absorbed the gluon propagators into the blob $F^{a_1 \cdots a_n}_{\mu_1\cdots \mu_n}$. Furthermore, we always define the blob as the sum of all possible crossings; it is thus symmetric under the simultaneous interchange of Lorentz, colour, and momentum indices. Because every Lorentz index of $F$ is contracted with the same vector $\hat{n}^\mu$, it is automatically symmetric in these. These two symmetries imply that an interchange of momentum variables is equivalent to an interchange of the corresponding colour indices. In particular, it is now straightforward to relate \eqref{eq: basicstructuresb} to \eqref{eq: basicstructuresa}:
\begin{align*}
  \begin{tikzfigure}{0.6}{0.75}
    \draw[gluon] (0.5,0) -- (0.5,-2);
    \draw[gluon] (1,0) -- (1,-2);
    \draw[gluon] (2.5,0) -- (2.5,-2);
    \node[gluontext] at (1.75,-0.75) {$\ldots$};
    \draw[wilson, wilsonarrowreversed] (0,0) -- (3,0);
    \filldraw[wilsontext] (3,0) circle(0.125);
    \filldraw[blob] (1.5,-2) circle(1.5 and 0.75);
    \node[blobtext] at (1.5,-2) {$F$};
  \end{tikzfigure}
% &\! = \;
%   \left(\i g\right)^n t^{a_n}\!\cdots t^{a_1}\! \Int{\mathcal{D}k}
%   I_n^\text{u.b.} \, F_{\mu_1\cdots\mu_n}^{a_1\cdots a_n}{ (k_1,\ldots, k_n)} \\
  &\! = \;
    \left(-\i g\right)^n t^{a_n}\!\cdots t^{a_1}\! \Int{\mathcal{D}k}
    I_n^\text{l.b.} \, F_{\mu_1\cdots\mu_n}^{a_n\cdots a_1}{ (k_1,\ldots, k_n)}.
\end{align*}
Often the blob has a factorable colour structure, i.e.\@
\begin{equation}
  F_{\mu_1\cdots\mu_n}^{a_1\cdots a_n} (k_1,\ldots, k_n) =
    C^{a_1\cdots a_n} F_{\mu_1\cdots\mu_n} (k_1,\ldots, k_n).
\end{equation}
If we now define the following notations:
\begin{subalign}
  C &= t^{a_n}\cdots t^{a_1} C^{a_1\cdots a_n}, &
  \bar{C} &= t^{a_n}\cdots t^{a_1} C^{a_n\cdots a_1},
\end{subalign}
we can simply write
\begin{subalign}
  \begin{tikzfigure}{0.6}{0.75}
    \draw[gluon] (0.5,0) -- (0.5,-2);
    \draw[gluon] (1,0) -- (1,-2);
    \draw[gluon] (2.5,0) -- (2.5,-2);
    \node[gluontext] at (1.75,-0.75) {$\ldots$};
    \draw[wilson, wilsonarrow] (0,0) -- (3,0);
    \filldraw[wilsontext] (0,0) circle(0.125);
    \filldraw[blob] (1.5,-2) circle(1.5 and 0.75);
    \node[blobtext] at (1.5,-2) {$F$};
  \end{tikzfigure}
  &\;= \; C
  \begin{tikzfigure}{0.6}{0.75}
    \draw[photon] (0.5,0) -- (0.5,-2);
    \draw[photon] (1,0) -- (1,-2);
    \draw[photon] (2.5,0) -- (2.5,-2);
    \node[photontext] at (1.75,-0.75) {$\ldots$};
    \draw[wilson, wilsonarrow] (0,0) -- (3,0);
    \filldraw[wilsontext] (0,0) circle(0.125);
    \filldraw[blob] (1.5,-2) circle(1.5 and 0.75);
    \node[blobtext] at (1.5,-2) {$F$};
  \end{tikzfigure},
  \\
  \begin{tikzfigure}{0.6}{0.75}
    \draw[gluon] (0.5,0) -- (0.5,-2);
    \draw[gluon] (1,0) -- (1,-2);
    \draw[gluon] (2.5,0) -- (2.5,-2);
    \node[gluontext] at (1.75,-0.75) {$\ldots$};
    \draw[wilson, wilsonarrowreversed] (0,0) -- (3,0);
    \filldraw[wilsontext] (3,0) circle(0.125);
    \filldraw[blob] (1.5,-2) circle(1.5 and 0.75);
    \node[blobtext] at (1.5,-2) {$F$};
  \end{tikzfigure}
  &\;= \; (-)^n \, \bar{C}
  \begin{tikzfigure}{0.6}{0.75}
    \draw[photon] (0.5,0) -- (0.5,-2);
    \draw[photon] (1,0) -- (1,-2);
    \draw[photon] (2.5,0) -- (2.5,-2);
    \node[photontext] at (1.75,-0.75) {$\ldots$};
    \draw[wilson, wilsonarrow] (0,0) -- (3,0);
    \filldraw[wilsontext] (0,0) circle(0.125);
    \filldraw[blob] (1.5,-2) circle(1.5 and 0.75);
    \node[blobtext] at (1.5,-2) {$F$};
  \end{tikzfigure}.
\end{subalign}
The yellow, `photon-like' wavy lines are just a reminder that there is no colour structure left in the blob. For a factorable blob example, take e.g.\@ the three gluon vertex:
\begin{equation*}
  F =
    g\, f^{a_1a_2a_3} \left[
      (k_1 - k_2)^\rho D_{\mu_1 \nu}(k_1) D^\nu_{\mu_2} (k_2) D_{\rho \mu_3}(k_3)
      +\text{cross.}
    \right],
\end{equation*}
with colour structure $f^{a_1 a_2 a_3}$. This implies that $\bar{C} = - C$:%, and hence
\begin{equation*}
  \begin{tikzfigure}{0.6}{0.25}
    \draw[gluon] (0.5,0) to[out=-90,in=-90,distance=40] (2.5,0);
    \draw[gluon] (1,0) -- (1.5,-1.2);
    \filldraw[gluontext] (1.5,-1.1) circle(0.1);
    \draw[wilson, wilsonarrowreversed] (0,0) -- (3,0);
    \filldraw[wilsontext] (3,0) circle(0.125);
  \end{tikzfigure}
  \quad = \quad
  \begin{tikzfigure}{0.6}{0.25}
    \draw[gluon] (0.5,0) to[out=-90,in=-90,distance=40] (2.5,0);
    \draw[gluon] (1,0) -- (1.5,-1.2);
    \filldraw[gluontext] (1.5,-1.1) circle(0.1);
    \draw[wilson, wilsonarrow] (0,0) -- (3,0);
    \filldraw[wilsontext] (0,0) circle(0.125);
  \end{tikzfigure}.
\end{equation*}
Of course, a lot of blob structures won't be colour factorable, but we can always write these as a sum of factorable terms:
\begin{equation*}
  F_{\mu_1\cdots\mu_n}^{a_1\cdots a_n} (k_1,\ldots, k_n) =
    \sum\limits_i C_i^{a_1\cdots a_n} F_{i\, \mu_1\cdots\mu_n} (k_1,\ldots, k_n),
\end{equation*}
such that we can repeat the same procedure as before
\begin{subalign}[eq: mixedcolourstructure]
  \begin{tikzfigure}{0.6}{0.75}
    \draw[gluon] (0.5,0) -- (0.5,-2);
    \draw[gluon] (1,0) -- (1,-2);
    \draw[gluon] (2.5,0) -- (2.5,-2);
    \node[gluontext] at (1.75,-0.75) {$\ldots$};
    \draw[wilson, wilsonarrow] (0,0) -- (3,0);
    \filldraw[wilsontext] (0,0) circle(0.125);
    \filldraw[blob] (1.5,-2) circle(1.5 and 0.75);
    \node[blobtext] at (1.5,-2) {$F$};
  \end{tikzfigure}
  &\;= \; \sum\limits_i C_i
  \begin{tikzfigure}{0.6}{0.75}
    \draw[photon] (0.5,0) -- (0.5,-2);
    \draw[photon] (1,0) -- (1,-2);
    \draw[photon] (2.5,0) -- (2.5,-2);
    \node[photontext] at (1.75,-0.75) {$\ldots$};
    \draw[wilson, wilsonarrow] (0,0) -- (3,0);
    \filldraw[wilsontext] (0,0) circle(0.125);
    \filldraw[blob] (1.5,-2) circle(1.5 and 0.75);
    \node[blobtext] at (1.5,-2) {$F_i$};
  \end{tikzfigure},
  \\
  \begin{tikzfigure}{0.6}{0.75}
    \draw[gluon] (0.5,0) -- (0.5,-2);
    \draw[gluon] (1,0) -- (1,-2);
    \draw[gluon] (2.5,0) -- (2.5,-2);
    \node[gluontext] at (1.75,-0.75) {$\ldots$};
    \draw[wilson, wilsonarrowreversed] (0,0) -- (3,0);
    \filldraw[wilsontext] (3,0) circle(0.125);
    \filldraw[blob] (1.5,-2) circle(1.5 and 0.75);
    \node[blobtext] at (1.5,-2) {$F$};
  \end{tikzfigure}
  &\;= \; (-)^n \, \sum\limits_i \bar{C}_i
  \begin{tikzfigure}{0.6}{0.75}
    \draw[photon] (0.5,0) -- (0.5,-2);
    \draw[photon] (1,0) -- (1,-2);
    \draw[photon] (2.5,0) -- (2.5,-2);
    \node[photontext] at (1.75,-0.75) {$\ldots$};
    \draw[wilson, wilsonarrow] (0,0) -- (3,0);
    \filldraw[wilsontext] (0,0) circle(0.125);
    \filldraw[blob] (1.5,-2) circle(1.5 and 0.75);
    \node[blobtext] at (1.5,-2) {$F_i$};
  \end{tikzfigure}.
\end{subalign}

\section{Piecewise Linear Wilson Lines}
When connecting a $n$-gluon blob to a piecewise Wilson line, the $n$ gluons aren't necessarily all connected to the same segment, but the $n$-gluons can be divided among several segments; this is the physical interpretation of formula \eqref{eq: PiecewiseIntegral}. Because the blob is summed over all crossings, multiple-segment terms can be related by straightforward substitution, e.g.
\begin{equation}\label{eq: multisegmentrelation}
  \mathcal{U}^J_1\mathcal{U}^K_2 =
    \mathcal{U}^J_2\mathcal{U}^K_1 \Big|_{ (r_J \leftrightarrow r_K, n_J \leftrightarrow n_K) },
\end{equation}
etc. When connecting e.g.\@  a $4$-gluon blob, we need to calculate exactly 5 diagrams, independent on the number of segments $M$. These diagrams are (cf.\@ \eqref{eq: PiecewiseIntegral}):
\begin{equation*}
  \mathcal{U}^J_4, \, \mathcal{U}^J_3\mathcal{U}^K_1, \, \mathcal{U}^J_2\mathcal{U}^K_2, \,
  \mathcal{U}^J_2\mathcal{U}^K_1\mathcal{U}^L_1, \, \text{and }
  \mathcal{U}^J_1\mathcal{U}^K_1\mathcal{U}^L_1\mathcal{U}^O_1.
\end{equation*}
They are the easiest represented schematically:
\begin{equation}\label{eq: 4gluondiags}
  \begin{tikzfigure}{0.55}{3.5}
    \draw[gluon] (0.75,0) -- (0.75,-2);
    \draw[gluon] (1.25,0) -- (1.25,-2);
    \draw[gluon] (1.75,0) -- (1.75,-2);
    \draw[gluon] (2.25,0) -- (2.25,-2);
    \draw[wilson, wilsonarrow] (0.25,0) -- (2.75,0);
    \filldraw[wilsontext] (0.25,0) circle(0.125);
    \filldraw[blob] (1.5,-2) circle(1.5 and 0.75);
    \node[blobtext] at (1.5,-2) {$F$};
    \begin{scope}[shift={(5,0)}]
      \draw[gluon] (0.5,0) -- (1.25,-2);
      \draw[gluon] (2.5,0) -- (1.75,-2);
      \draw[gluon] (3,0) -- (2.125,-2);
      \draw[gluon] (3.5,0) -- (2.5,-2);
      \draw[wilson, wilsonarrow] (-0.25,0) -- (1.25,0);
      \draw[wilson, wilsonarrow] (2,0) -- (4,0);
      \filldraw[wilsontext] (-0.25,0) circle(0.125);
      \filldraw[wilsontext] (2,0) circle(0.125);
      \filldraw[blob] (1.5,-2) circle(1.5 and 0.75);
      \node[blobtext] at (1.5,-2) {$F$};
    \end{scope}
    \begin{scope}[shift={(11,0)}]
      \draw[gluon] (0,0) -- (0.75,-2);
      \draw[gluon] (0.5,0) -- (1.25,-2);
      \draw[gluon] (2.5,0) -- (1.75,-2);
      \draw[gluon] (3,0) -- (2.25,-2);
      \draw[wilson, wilsonarrow] (-0.5,0) -- (1,0);
      \draw[wilson, wilsonarrow] (2,0) -- (3.5,0);
      \filldraw[wilsontext] (-0.5,0) circle(0.125);
      \filldraw[wilsontext] (2,0) circle(0.125);
      \filldraw[blob] (1.5,-2) circle(1.5 and 0.75);
      \node[blobtext] at (1.5,-2) {$F$};
    \end{scope}
    \begin{scope}[shift={(1,-4)}]
      \draw[gluon] (1.5,0) -- (1.5,-2);
      \draw[gluon] (3,0) -- (2,-2);
      \draw[gluon] (3.5,0) -- (2.5,-2);
      \draw[gluon] (0,0) -- (1,-2);
      \draw[wilson, wilsonarrow] (0.75,0) -- (2.25,0);
      \draw[wilson, wilsonarrow] (-1,0) -- (0.5,0);
      \draw[wilson, wilsonarrow] (2.5,0) -- (4,0);
      \filldraw[wilsontext] (-1,0) circle(0.125);
      \filldraw[wilsontext] (2.5,0) circle(0.125);
      \filldraw[wilsontext] (0.75,0) circle(0.125);
      \filldraw[blob] (1.5,-2) circle(1.5 and 0.75);
      \node[blobtext] at (1.5,-2) {$F$};
    \end{scope}
    \begin{scope}[shift={(8.5,-4)}]
      \draw[gluon] (0.5,0) -- (1.5,-2);
      \draw[gluon] (-1,0) -- (1,-2);
      \draw[gluon] (2.5,0) -- (1.5,-2);
      \draw[gluon] (4,0) -- (2,-2);
      \draw[wilson, wilsonarrow] (0,0) -- (1.25,0);
      \draw[wilson, wilsonarrow] (-1.5,0) -- (-0.25,0);
      \draw[wilson, wilsonarrow] (1.75,0) -- (3,0);
      \draw[wilson, wilsonarrow] (3.25,0) -- (4.5,0);
      \filldraw[wilsontext] (0,0) circle(0.125);
      \filldraw[wilsontext] (-1.5,0) circle(0.125);
      \filldraw[wilsontext] (1.75,0) circle(0.125);
      \filldraw[wilsontext] (3.25,0) circle(0.125);
      \filldraw[blob] (1.5,-2) circle(1.5 and 0.75);
      \node[blobtext] at (1.5,-2) {$F$};
    \end{scope}
  \end{tikzfigure}
\end{equation}
% In addition, there are 3 more diagrams that can be related using \eqref{eq: multisegmentrelation}, built from the segments
% \begin{equation*}
%   \mathcal{U}^J_1\mathcal{U}^K_3, \,
%   \mathcal{U}^J_1\mathcal{U}^K_2\mathcal{U}^L_1, \, \text{and }
%   \mathcal{U}^J_1\mathcal{U}^K_1\mathcal{U}^L_2.
% \end{equation*}

Now what about the different path structures, as defined in \eqref{eq: basicstructures}? We can use the same trick as in the end of the former section, viz.\@ a sign change and an interchange of the corresponding colour indices. For instance:
\begin{equation*}
  \begin{tikzfigure}{0.6}{0.75}
    \draw[gluon] (0,0) -- (0.75,-2);
    \draw[gluon] (0.5,0) -- (1.25,-2);
    \draw[gluon] (2.5,0) -- (1.75,-2);
    \draw[gluon] (3,0) -- (2.25,-2);
    \draw[wilson, wilsonarrow] (-0.5,0) -- (1,0);
    \draw[wilson, wilsonarrowreversed] (2,0) -- (3.5,0);
    \filldraw[wilsontext] (-0.5,0) circle(0.125);
    \filldraw[wilsontext] (3.5,0) circle(0.125);
    \filldraw[blob] (1.5,-2) circle(1.5 and 0.75);
    \node[blobtext] at (1.5,-2) {$F$};
  \end{tikzfigure}
  \quad = \quad
  (-)^2
  \begin{tikzfigure}{0.6}{0.75}
    \draw[gluon] (0,0) -- (0.75,-2);
    \draw[gluon] (0.5,0) -- (1.25,-2);
    \draw[gluon] (2.5,0) -- (1.75,-2);
    \draw[gluon] (3,0) -- (2.25,-2);
    \draw[wilson, wilsonarrow] (-0.5,0) -- (1,0);
    \draw[wilson, wilsonarrow] (2,0) -- (3.5,0);
    \filldraw[wilsontext] (-0.5,0) circle(0.125);
    \filldraw[wilsontext] (2,0) circle(0.125);
    \filldraw[blob] (1.5,-2) circle(1.5 and 0.75);
    \node[blobtext] at (1.5,-2) {$ F \big|_{a_3\leftrightarrow a_4}$};
  \end{tikzfigure}.
\end{equation*}
The easiest way to implement this, is to define a path function $\Phi$ per diagram for a given blob, that gives the colour structure in function of the path type. For the leading order 2 gluon blob, this is straightforward:
\begin{subalign}[eq: 2gluonpathconstants]
  \begin{tikzfigure}{0.6}{0.2}
    \draw[gluon] (0.25,0) to[out=-60,in=-120,distance=25] (2,0);
    \draw[wilson,wilsonarrow] (-0.25,0) -- (2.25,0);
    \filldraw[wilsontext] (-0.25,0) circle(0.125);
  \end{tikzfigure}
  &:
  \qquad \Phi(J) = C_F,
  \\
  \begin{tikzfigure}{0.6}{0.2}
    \draw[gluon] (0.5,0) to[out=-90,in=-90,distance=30] (2.5,0);
    \draw[wilson,wilsonarrow] (-0.5,0) -- (1,0);
    \filldraw[wilsontext] (-0.5,0) circle(0.125);
    \draw[wilson,wilsonarrow] (1.5,0) -- (3,0);
    \filldraw[wilsontext] (1.5,0) circle(0.125);
  \end{tikzfigure}
  &:
  \qquad \Phi(J,K) = (-)^{\phi_J + \phi_K}C_F,
\end{subalign}
where $\phi_J$ represents the structure of the segment:
\begin{equation}
  \phi_J = \begin{cases}
    0 & J =
    \begin{tikzfigure}{0.5}{0.12}
      \draw[wilson,wilsonarrow] (-0.25,0) -- (1.25,0);
      \filldraw[wilsontext] (-0.25,0) circle(0.125);
    \end{tikzfigure} \\
    1 & J =
    \begin{tikzfigure}{0.5}{0.12}
      \draw[wilson,wilsonarrowreversed] (-0.25,0) -- (1.25,0);
      \filldraw[wilsontext] (1.25,0) circle(0.125);
    \end{tikzfigure}
  \end{cases}
\end{equation}
Keep in mind that in our original definition of the Wilson line \eqref{eq: Wilsondef}, colour indices are not yet traced, hence equations \eqref{eq: 2gluonpathconstants} should still be multiplied with an unit matrix $\1$. Similarly for the leading order 3-gluon blob we find:
\begin{subalign}[eq: 3gluonpathconstants]
  \begin{tikzfigure}{0.6}{0.2}
    \draw[gluon] (0,0) to[out=-60,in=-120,distance=30] (2,0);
    \draw[gluon] (1,0) -- (1,-0.7);
    \draw[wilson,wilsonarrow] (-0.25,0) -- (2.25,0);
    \filldraw[wilsontext] (-0.25,0) circle(0.125);
    \filldraw[gluontext] (1,-0.7) circle(0.125);
  \end{tikzfigure}
  &: \,
  \Phi(J) = -\i \frac{N}{2}C_F,
  \\
  \begin{tikzfigure}{0.6}{0.2}
    \draw[gluon] (0.5,0) to[out=-90,in=-90,distance=30] (2.5,0);
    \draw[gluon] (2,0) -- (1.5,-0.75);
    \draw[wilson,wilsonarrow] (-0.5,0) -- (1,0);
    \filldraw[wilsontext] (-0.5,0) circle(0.125);
    \draw[wilson,wilsonarrow] (1.5,0) -- (3,0);
    \filldraw[wilsontext] (1.5,0) circle(0.125);
    \filldraw[gluontext] (1.5,-0.75) circle(0.125);
  \end{tikzfigure}
  &: \,
  \Phi(J,K) = -\i(-)^{\phi_J+\phi_K} \frac{N}{2}C_F,
  \\
  \begin{tikzfigure}{0.6}{0.2}
    \draw[gluon] (1,0) to[out=-90,in=-90,distance=27] (4,0);
    \draw[gluon] (2.5,0) -- (2.5,-0.75);
    \draw[wilson,wilsonarrow] (0.5,0) -- (1.5,0);
    \filldraw[wilsontext] (0.5,0) circle(0.125);
    \draw[wilson,wilsonarrow] (2,0) -- (3,0);
    \filldraw[wilsontext] (2,0) circle(0.125);
    \draw[wilson,wilsonarrow] (3.5,0) -- (4.5,0);
    \filldraw[wilsontext] (3.5,0) circle(0.125);
    \filldraw[gluontext] (2.5,-0.75) circle(0.125);
  \end{tikzfigure}
  &: \,
  \Phi(\ldots) = -\i (-)^{\phi_J+\phi_K+\phi_L}\frac{N}{2}C_F.
\end{subalign}
For non-factorable blobs we use the same trick as in \eqref{eq: mixedcolourstructure}, by giving $\Phi$ an extra index to identify the sub diagram it belongs to.

Let us introduce a new notation, to indicate a full diagram, but without the colour content, in which a blob is connected to $m$ Wilson line segments, with $n_i$ gluons connected to the $i$-th segment:
\begin{equation}
  \mathcal{W}_{n_m\cdots n_1}^{J_m\cdots J_1},
\end{equation}
where we will write the indices from right to left for convenience.
Returning to the 4-gluon blob, we can now write the full result for a factorable blob using \eqref{eq: PiecewiseIntegral}:
\begin{multline}\label{eq: 4gluonblobresult}
  \mathcal{U}^4 = 
    \sum_J^M \Phi_4\mathcal{W}_{4}^{J} +
    \sum_{J=2}^M \sum_{K=1}^{J-1}
      \left[ \Phi_{3\,1} \mathcal{W}_{3\,1}^{JK} + \Phi_{2\,2} \mathcal{W}_{2\,2}^{JK} \right] \\
    +
    \sum_{J=3}^M \sum_{K=2}^{J-1} \sum_{L=1}^{K-1} \Phi_{2\,1\,1} \mathcal{W}_{2\,1\,1}^{JKL}\\
    +
    \sum_{J=4}^M \sum_{K=3}^{J-1} \sum_{L=2}^{K-1} \sum_{O=1}^{L-1} 
      \Phi_{1\,1\,1\,1} \mathcal{W}_{1\,1\,1\,1}^{JKLO} +
    \text{symm.},
\end{multline}
where the symmetrised diagrams $\Phi_{1\,3} \mathcal{W}_{1\,3}^{JK}$, $\Phi_{1\,2\, 1} \mathcal{W}_{1\,2\,1}^{JKL}$, and $\Phi_{1\,1\,2} \mathcal{W}_{1\,1\,2}^{JKL}$ are calculated using \eqref{eq: multisegmentrelation}, interchanging also the $\phi_J$.
%In other words:
% \begin{subalign}
%   \Phi_{1\,3}(J,K) \, \mathcal{W}_{1\,3}^{JK} &=
%     \Phi_{3\,1}(K,J) \, \mathcal{W}_{3\,1}^{KJ} \\
%   \Phi_{1\,2\, 1}(J,K,L) \, \mathcal{W}_{1\,2\,1}^{JKL} &=
%     \Phi_{2\,1\, 1}(J,L,K) \, \mathcal{W}_{2\,1\,1}^{JLK} \\
%   \Phi_{1\,1\,2}(J,K,L) \, \mathcal{W}_{1\,1\,2}^{JKL} &=
%     \Phi_{2\,1\, 1}(L,K,J) \, \mathcal{W}_{2\,1\,1}^{LKJ}
%\end{subalign}
For a non-factorable blob, every term is just replaced by a sum over sub diagrams.
% , e.g.
% \begin{equation}
%   \Phi_4\mathcal{W}_{4}^{J} \; \rightarrow \; \sum_i \Phi_4^i \mathcal{W}_{4}^{i\, J}.
% \end{equation}
It is important to realise that both the $\Phi_{n_i\cdots}$ and $\mathcal{W}_{n_i \cdots}$ can be calculated independent of the path structure, giving a result depending on $n_J$, $r_J$ and $\phi_J$.
\\

So far we have only calculated amplitudes. To get probabilities from these, we can do this in the standard way, viz.\@ squaring diagrams and combining them order by order (squared terms and interference terms), or we could treat the squared diagram as one Wilson line, where the segments to the right of the cut are the hermitian conjugate of those to the left . We choose to continue with the latter case, where we now have three distinct sectors of diagrams: a sector where the blob is only connecting segments left of the cut, a sector where the blob is only connecting segments right of the cut, and a sector where the blob is connecting segments both left and right of the cut. In other words:
\begin{equation}
  \mathcal{U} = \mathcal{U}_\text{left} + \mathcal{U}_\text{cut} + \mathcal{U}_\text{right}.
\end{equation}
For the first two nothing changes, the calculations go as before. For the example of the 4-gluon blob, the first sector $\mathcal{U}^4_\text{left}$ is almost exactly equal to \eqref{eq: 4gluonblobresult}, but the sums run up only to $M_c$, the number of segments before the cut, instead of $M$. The last sector $\mathcal{U}^4_\text{right}$ is simply the hermitian conjugate of this, starting at $M_c+1$.
For the remaining sector $\mathcal{W}_\text{cut}$ we need to define a cut blob. Given a blob, several possible cut blobs might exist, depending on the number of gluons to the left and right of the cut. E.g.\@ the leading order 4-gluon cut blobs are given by
\begin{subalign}
  \begin{tikzfigure}{0.5}{1} \label{eq: 4gluoncutblob13}
    \draw[gluon] (0.3,-1) -- (1,-2.2);
    \draw[gluon] (1.7,-1) -- (1.2,-2.2);
    \draw[gluon] (2.2,-1) -- (1.7,-2.2);
    \draw[gluon] (2.7,-1) -- (2,-2.2);
    \filldraw[blob] (1.5,-2) circle(1. and 0.5);
    \draw[finalstatecut] (1.2,-2.75) -- (1.2,-0.9);
  \end{tikzfigure}
  &=
  \begin{tikzfigure}{0.5}{0.25}
    \draw[gluon] (0,0) to[out=-60, in=-120,distance=25] (1.5,0);
    \draw[gluon] (2,0) to[out=-60, in=-120,distance=25] (3.5,0);
    \draw[finalstatecut] (0.75,0.2) -- (0.75,-1);
  \end{tikzfigure}
  +\text{cross.},
  \\
  \begin{tikzfigure}{0.5}{1} \label{eq: 4gluoncutblob22}
    \draw[gluon] (0.3,-1) -- (1,-2.2);
    \draw[gluon] (1,-1) -- (1.2,-2.2);
    \draw[gluon] (2,-1) -- (1.7,-2.2);
    \draw[gluon] (2.7,-1) -- (2,-2.2);
    \filldraw[blob] (1.5,-2) circle(1. and 0.5);
    \draw[finalstatecut] (1.5,-2.75) -- (1.5,-0.9);
  \end{tikzfigure}
  &=
  \begin{tikzfigure}{0.5}{0.25}
    \draw[gluon] (0,0) to[out=-60, in=-120,distance=25] (1.5,0);
    \draw[gluon] (2,0) to[out=-60, in=-120,distance=25] (3.5,0);
    \draw[finalstatecut] (1.75,0.25) -- (1.75,-1);
  \end{tikzfigure}
  +
  \begin{tikzfigure}{0.5}{0.4}
    \draw[gluon] (0,0) to[out=-60, in=-120,distance=42] (2.5,0);
    \draw[gluon] (0.5,0) to[out=-60, in=-120,distance=25] (2,0);
    \draw[finalstatecut] (1.25,0.) -- (1.25,-1.5);
  \end{tikzfigure}
  +\text{cross.},
\end{subalign}
where the crossings are to be made on the sides of the cut separately. Also note that when the cut blob is more complex, it should be summed over all possible cut locations. Consider e.g.\@ the fermionic part of the NLO 2-gluon cut blob:
\begin{equation}
  \begin{tikzfigure}{0.5}{1}
    \draw[gluon] (0.3,-1) -- (1,-2.2);
    \draw[gluon] (2.7,-1) -- (2,-2.2);
    \filldraw[blob] (1.5,-2) circle(1. and 0.5);
    \draw[finalstatecut] (1.5,-2.75) -- (1.5,-1.25);
  \end{tikzfigure}
  =
  \begin{tikzfigure}{0.5}{0.1}
    \draw[gluon] (0,0) -- (1.,0);
    \draw[gluon] (2.25,0) -- (3.25,0);
    \draw[quark] (1.,0) to[out=80,in=100] (2.25,0);
    \draw[quark] (2.25,0) to[out=-100,in=-80] (1.,0);
    \draw[finalstatecut] (0.5,-0.75) -- (0.5,0.75);
  \end{tikzfigure}
  +
  \begin{tikzfigure}{0.5}{0.1}
    \draw[gluon] (0,0) -- (1.,0);
    \draw[gluon] (2.25,0) -- (3.25,0);
    \draw[quark] (1.,0) to[out=80,in=100] (2.25,0);
    \draw[quark] (2.25,0) to[out=-100,in=-80] (1.,0);
    \draw[finalstatecut] (1.45,-0.75) -- (1.45,0.75);
  \end{tikzfigure}
  +
  \begin{tikzfigure}{0.5}{0.1}
    \draw[gluon] (0,0) -- (1.,0);
    \draw[gluon] (2.25,0) -- (3.25,0);
    \draw[quark] (1.,0) to[out=80,in=100] (2.25,0);
    \draw[quark] (2.25,0) to[out=-100,in=-80] (1.,0);
    \draw[finalstatecut] (2.75,-0.75) -- (2.75,0.75);
  \end{tikzfigure}.
\end{equation}
Returning to the sector $\mathcal{W}_\text{cut}$, we investigate which diagrams are added in comparison to \eqref{eq: 4gluondiags} due to the cut. First note that a Wilson line segment is never cut itself by its nature, so cuts are always placed \emph{between} two segments.\footnote{There seem to exist exceptions to this in existing literature, but it is merely a matter of naming conventions. E.g.\@ in \cite{Collins:2011zzd} the definition of a collinear PDF (a cut diagram itself) is made gauge-invariant by adding a finite Wilson line that is cut. The exponential coming from \eqref{eq: FRextpoint} is integrated over $a^\mu$, giving a delta function, which in turn can be associated with \eqref{eq: FRcutprop}. Hence the identification as a cut Wilson line.} Another remark is that we cannot simply use relation \eqref{eq: multisegmentrelation} as before, because it could change the cut topology. Cut diagrams are sorted depending on how its gluons are distributed on the left resp.\@ right side of the cut, and connected to the appropriate cut blob. For instance the second diagram of \eqref{eq: 4gluondiags}, $\mathcal{W}^4_{3\,1}$ can be cut in one way only, connecting the Wilson line to cut blob \eqref{eq: 4gluoncutblob13}, but the fourth diagram, $\mathcal{W}^4_{2\,1\,1}$, can be cut in two ways: cutting with one gluon on the left (written as $\mathcal{W}^4_{2\,1|1}$ and connected to \eqref{eq: 4gluoncutblob13}), or cutting with two gluons (written as $\mathcal{W}^4_{2|1\,1}$ and connected to \eqref{eq: 4gluoncutblob22}). Other cut topologies can be related by Hermitian conjugation when switching left and right sides, e.g.\@ $
  \mathcal{W}^{JKL}_{1\,1|2} = \mathcal{W}^{\dagger\, LKJ}_{2|1\,1}$.
In case of the 4-gluon blob, the following diagrams have to be added to \eqref{eq: 4gluondiags}:
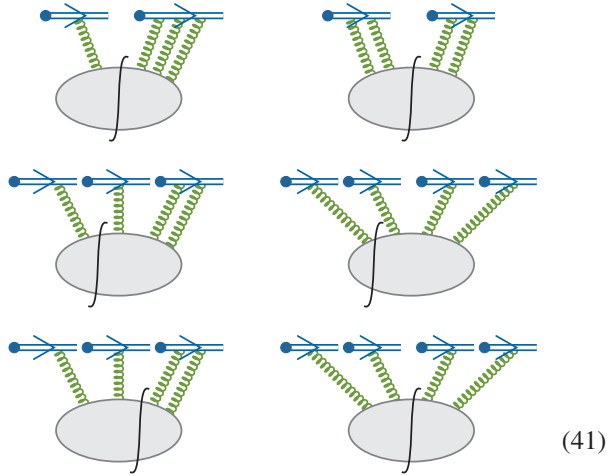
\begin{equation}
  \begin{tikzfigure}{0.55}{5.75}
    \begin{scope}[shift={(1,0)}]
      \draw[gluon] (0.5,0) -- (1.25,-2);
      \draw[gluon] (2.5,0) -- (1.75,-2);
      \draw[gluon] (3,0) -- (2.125,-2);
      \draw[gluon] (3.5,0) -- (2.5,-2);
      \draw[wilson, wilsonarrow] (-0.25,0) -- (1.25,0);
      \draw[wilson, wilsonarrow] (2,0) -- (4,0);
      \filldraw[wilsontext] (-0.25,0) circle(0.125);
      \filldraw[wilsontext] (2,0) circle(0.125);
      \filldraw[blob] (1.5,-2) circle(1.5 and 0.75);
      \draw[finalstatecut] (1.5,-3) -- (1.5,-1);
    \end{scope}
    \begin{scope}[shift={(8,0)}]
      \draw[gluon] (0,0) -- (0.75,-2);
      \draw[gluon] (0.5,0) -- (1.25,-2);
      \draw[gluon] (2.5,0) -- (1.75,-2);
      \draw[gluon] (3,0) -- (2.25,-2);
      \draw[wilson, wilsonarrow] (-0.5,0) -- (1,0);
      \draw[wilson, wilsonarrow] (2,0) -- (3.5,0);
      \filldraw[wilsontext] (-0.5,0) circle(0.125);
      \filldraw[wilsontext] (2,0) circle(0.125);
      \filldraw[blob] (1.5,-2) circle(1.5 and 0.75);
      \draw[finalstatecut] (1.5,-3) -- (1.5,-1);
    \end{scope}
    \begin{scope}[shift={(1,-4)}]
      \draw[gluon] (1.5,0) -- (1.5,-2);
      \draw[gluon] (3,0) -- (2,-2);
      \draw[gluon] (3.5,0) -- (2.5,-2);
      \draw[gluon] (0,0) -- (1,-2);
      \draw[wilson, wilsonarrow] (0.75,0) -- (2.25,0);
      \draw[wilson, wilsonarrow] (-1,0) -- (0.5,0);
      \draw[wilson, wilsonarrow] (2.5,0) -- (4,0);
      \filldraw[wilsontext] (-1,0) circle(0.125);
      \filldraw[wilsontext] (2.5,0) circle(0.125);
      \filldraw[wilsontext] (0.75,0) circle(0.125);
      \filldraw[blob] (1.5,-2) circle(1.5 and 0.75);
      \draw[finalstatecut] (1,-3) -- (1,-1);
    \end{scope}
    \begin{scope}[shift={(1,-8)}]
      \draw[gluon] (1.5,0) -- (1.5,-2);
      \draw[gluon] (3,0) -- (2,-2);
      \draw[gluon] (3.5,0) -- (2.5,-2);
      \draw[gluon] (0,0) -- (1,-2);
      \draw[wilson, wilsonarrow] (0.75,0) -- (2.25,0);
      \draw[wilson, wilsonarrow] (-1,0) -- (0.5,0);
      \draw[wilson, wilsonarrow] (2.5,0) -- (4,0);
      \filldraw[wilsontext] (-1,0) circle(0.125);
      \filldraw[wilsontext] (2.5,0) circle(0.125);
      \filldraw[wilsontext] (0.75,0) circle(0.125);
      \filldraw[blob] (1.5,-2) circle(1.5 and 0.75);
      \draw[finalstatecut] (2,-3) -- (2,-1);
    \end{scope}
    \begin{scope}[shift={(8,-4)}]
      \draw[gluon] (0.5,0) -- (1.5,-2);
      \draw[gluon] (-1,0) -- (1,-2);
      \draw[gluon] (2.5,0) -- (1.5,-2);
      \draw[gluon] (4,0) -- (2,-2);
      \draw[wilson, wilsonarrow] (0,0) -- (1.25,0);
      \draw[wilson, wilsonarrow] (-1.5,0) -- (-0.25,0);
      \draw[wilson, wilsonarrow] (1.75,0) -- (3,0);
      \draw[wilson, wilsonarrow] (3.25,0) -- (4.5,0);
      \filldraw[wilsontext] (0,0) circle(0.125);
      \filldraw[wilsontext] (-1.5,0) circle(0.125);
      \filldraw[wilsontext] (1.75,0) circle(0.125);
      \filldraw[wilsontext] (3.25,0) circle(0.125);
      \filldraw[blob] (1.5,-2) circle(1.5 and 0.75);
      \draw[finalstatecut] (0.6,-3) -- (0.6,-1);
    \end{scope}
    \begin{scope}[shift={(8,-8)}]
      \draw[gluon] (0.5,0) -- (1.5,-2);
      \draw[gluon] (-1,0) -- (1,-2);
      \draw[gluon] (2.5,0) -- (1.5,-2);
      \draw[gluon] (4,0) -- (2,-2);
      \draw[wilson, wilsonarrow] (0,0) -- (1.25,0);
      \draw[wilson, wilsonarrow] (-1.5,0) -- (-0.25,0);
      \draw[wilson, wilsonarrow] (1.75,0) -- (3,0);
      \draw[wilson, wilsonarrow] (3.25,0) -- (4.5,0);
      \filldraw[wilsontext] (0,0) circle(0.125);
      \filldraw[wilsontext] (-1.5,0) circle(0.125);
      \filldraw[wilsontext] (1.75,0) circle(0.125);
      \filldraw[wilsontext] (3.25,0) circle(0.125);
      \filldraw[blob] (1.5,-2) circle(1.5 and 0.75);
      \draw[finalstatecut] (1.5,-3) -- (1.5,-1);
    \end{scope}
  \end{tikzfigure}
\end{equation}
Now  we have the necessary ingredients to write the cut sector for the 4-gluon blob:
\begin{multline}
  \mathcal{W}^4_\text{cut} =
    \sum_{M_c+1}^M \sum_1^{M_c}\left[
      \left(\mathcal{W}_{3|1}^{JK}+\text{h.c.}\right) + \mathcal{W}_{2|2}^{JK}
    \right] \\
    +\sum_{M_c+ 2}^M \sum_{M_c+1}^{J-1} \sum_1^{M_c}\left[
      \left( \mathcal{W}_{2\,1|1}^{JKL}+\text{symm.}\right)
      + \mathcal{W}_{2|1\,1}^{JKL}
    \right] \\
    +\sum_{M_c+ 2}^M \sum_{M_c+1}^{J-1}\sum_{2}^{M_c} \sum_1^{L-1} \mathcal{W}_{1\,1|1\,1}^{JKLO}
    \\
    +\sum_{M_c+ 3}^M \sum_{M_c+2}^{J-1}\sum_{M_c+1}^{K-1} \sum_1^{M_c} \mathcal{W}_{1\,1\,1|1}^{JKLO} +\text{h.c.}
\end{multline}

\section{Example Calculation}
Let us recapitulate our framework with a small example; we will calculate the 2-gluon blob connected to a general Wilson line at NLO. At any order, there are 3 possible 2-gluon diagrams:
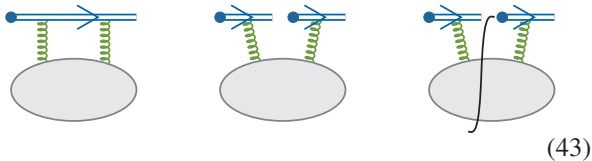
\begin{equation}
\begin{tikzfigure}{0.55}{0.8}
  \draw[gluon] (0.75,0) -- (0.75,-1.75);
  \draw[gluon] (2.25,0) -- (2.25,-1.75);
  \filldraw[blob] (1.5,-1.75) circle(1.5 and 0.75);
  \draw[wilson,wilsonarrow] (0,0) -- (3,0);
  \filldraw[wilsontext] (0,0) circle(0.125);
  \begin{scope}[shift={(5,0)}]
    \draw[gluon] (0.625,0) -- (1,-1.75);
    \draw[gluon] (2.325,0) -- (2,-1.75);
    \filldraw[blob] (1.5,-1.75) circle(1.5 and 0.75);
    \draw[wilson,wilsonarrow] (0,0) -- (1.25,0);
    \filldraw[wilsontext] (0,0) circle(0.125);
    \draw[wilson,wilsonarrow] (1.75,0) -- (3,0);
    \filldraw[wilsontext] (1.75,0) circle(0.125);
  \end{scope}
  \begin{scope}[shift={(10,0)}]
    \draw[gluon] (0.625,0) -- (1,-1.75);
    \draw[gluon] (2.325,0) -- (2,-1.75);
    \filldraw[blob] (1.5,-1.75) circle(1.5 and 0.75);
    \draw[wilson,wilsonarrow] (0,0) -- (1.25,0);
    \filldraw[wilsontext] (0,0) circle(0.125);
    \draw[wilson,wilsonarrow] (1.75,0) -- (3,0);
    \filldraw[wilsontext] (1.75,0) circle(0.125);
    \draw[finalstatecut] (1.25,-2.75) -- (1.25,0);
  \end{scope},
\end{tikzfigure}
\end{equation}
At NLO the blob is just a LO gluon propagator (here given in Feynman gauge):
\begin{equation}
  F = -\i (2\pi)^4 \diracdelta{k_1+k_2} \delta^{ab} \frac{1}{k_1^2+\i\varepsilon}g^{\mu_1\mu_2},
\end{equation}
and the cut blob just a radiated gluon integrated over:
\begin{equation}
  F = - (2\pi)^5 \diracdelta{k_1+k_2}\delta^{ab} \heavisidetheta{k_1^+}\diracdelta{k_1^2} g^{\mu\nu}.
\end{equation}
The colour factors (for both) are given by \eqref{eq: 2gluonpathconstants}. The first diagram is then given by, using dimensional regularisation:
\begin{align}
  \mathcal{W}_2^J &=
    -\i \,g^2 C_F \mu^{2 \epsilon} \, \hat{n}_J^2
    \Int{\frac{\Dif[\omega]{k}}{\left(2\pi\right)^\omega}}
    \frac{1}{\hat{n}_J \!\cdot\! k + \i\eta} \frac{1}{\i\eta}
    \frac{1}{k_1^2+\i\varepsilon} \\
  &=
    -\frac{1}{2}\alpha_s C_F \left(1 + \epsilon \ln 4\pi\mu^2\frac{\hat{n}_J^2}{\eta^2}\right)
\end{align}

\section*{Acknowledgements}
I am very grateful to the organisers of the Transversity 2014 conference for creating such a fruitful environment. Furthermore I would like to thank I.O.~Cherednikov, M.~Echevarria, L.~Gamberg, A.~Idilbi,T.~Mertens, A.~Prokudin and P.~Taels for useful discussions and insights.

\nocite{*}
\bibliographystyle{woc}
\bibliography{Bibliography}

\end{document}